\documentstyle[12pt]{article}

\begin{document}
\baselineskip1.6\baselineskip
\parindent1.0\parindent

\begin{titlepage}
\begin{center}

{\large{\bf Black Holes in Two Dimensional Dilaton Gravity and\\
 Nonlinear Klein-Gordon Soliton}}

\vspace{1.5cm}

Hyeon-Min Johng, Hak-Soo Shin\footnote{E-Mail:hsshin@ibm3090.snu.ac.kr}
 and Kwang-Sup Soh\footnote{E-Mail:kssoh@phyb.snu.ac.kr}
\\
\em{Department of Physics Education,
Seoul National University,\\ Seoul 151-742, Korea\\}
\end{center}
\vspace{4cm}

\begin{abstract}
\noindent
Two-dimensional dilaton gravity coupled to a Klein-Gordon
matter field with a quartic interaction term is considered.
The theory has a classical solution which exhibits black hole
formation by a soliton. The geometry of black hole induced
by a soliton is investigated.
\end{abstract}
\end{titlepage}

\pagestyle{plain}
\section{Introduction}
\hspace{0.5cm}
Black hole solutons of the two-dimensional dilaton gravity
 coupled with massless scalar field have attracted
 much interest since the pioneering work by Callan, Giddings,
 Harvey and Strominger(CGHS).$^1$ It was hoped that the
 model would unravel mysteries concerning the black hole
 evaporation and the resulting information loss. The theory, however,
 turned out to be intractable even in semi-classical
 approximations $^{2,3,4}$ let alone in full
 quantum analysis.

Much of the analysis of two dimensional models about black
 hole formation was only concerned with free scalar fields.$^1$
 Recently black hole formation by an interacting scalar field was
 considered,$^{5,6}$ where soliton solutions of Sine-Gordon
 theory provide energy-momentum stress tensor to cause the black hole. In
 two-dimensional spacetime there have been many types of partial
 differential equations besides Sine-Gordon theory which have soliton
 solutions. In particular Klein-Gordon theory with a quartic interaction
 offers another simple example of solitons.$^7$ In this paper
 we consider black hole formation in two-dimensional dilaton gravity
 coupled to a Klein-Gordon matter field with a quartic interaction term.

In section 2 we introduce the model by giving the action, choose
 the convenient gauge, i.e., conformal gauge, and further reduce the
 equations by making use of subgauge freedom. In section 3 we solve
 the equations exactly using the known soliton solutions of quartic
 Klein-Gordon theory, and study the geometry with an emphasis on
 black hole formation. In the last section relativistic limit of our
 solution is compared with the shock wave solution of the CGHS model.$^1$
\vspace{1cm}

\section{Action and Field Equations}
\hspace{0.5cm}
The action of the model we are concerned is
\begin{equation}
S=\frac{1}{2\pi}\int_M d^2 x \sqrt{-g}\left[e^{-2\phi}
\left(R+4(\nabla \phi)^2 +4\lambda^2\right)-
\frac{1}{2}(\nabla f)^2-\frac{\beta}{4}\left(f^2-\frac{\mu^2}
{\beta}\right)^2 e^{-2\phi}\right], \label{eq1}
\end{equation}
where $g$, $\phi$ and $f$ are metric, dilaton, and matter fields,
 respectively. $R$ is the curvature scalar, $\lambda^2$ is a cosmological
 constant, $\beta$ is a coupling constant of the matter field,
 and $\pm\sqrt{\frac{\mu^2}{\beta}}$
 is the minimum position of the quartic potential. Note that this is the
 CGHS action $^1$ except the last quartic interaction term which
 is introduced in order to study formation of black holes by the solitons.

The classical theory described by (\ref{eq1}) is most easily analyzed in the
conformal gauge:
\begin{eqnarray}
ds^2 &=& -e^{2\rho}dx^+dx^-, \label{eq2}
\\
g_{+-}& =& -\frac{1}{2}e^{2\rho},~~~~~ g^{+-} = -2e^{-2\rho},\label{eq3}
\\
R& =& 8e^{-2\rho}\partial_+\partial_-\rho,\label{eq4}
\end{eqnarray}
where $x^{\pm}=t \pm x$.
The action then reduces to
\begin{equation}
S = \frac{1}{\pi} \int d^2x \left[e^{-2\phi}
(2\partial_+\partial_-\rho -4\partial_+\phi\partial_-\phi
+\lambda^2 e^{2\rho})+\frac{1}{2}\partial_+f\partial_-f
-\frac{\beta}{16}(f^2-\frac{\mu^2}{\beta})^2 e^{2\rho-2\phi}\right],
\label{eq5}
\end{equation}
and the metric equations of motion are
\begin{eqnarray}
T_{++} &=& e^{-2\phi} (4\partial_+\rho\partial_+\phi -2\partial_+^2\phi)+
\frac{1}{2}(\partial_+f)^2 =0, \label{eq6}
\\
T_{--} &=&e^{-2\phi}(4\partial_-\rho\partial_-\phi -2\partial_-^2\phi)
+ \frac{1}{2}(\partial_-f)^2 =0,   \label{eq7}
\\
T_{+-}&=& e^{-2\phi}(2\partial_+\partial_-\phi -
4\partial_+\phi\partial_-\phi-\lambda^2e^{2\rho})
+\frac{\beta}{16}(f^2-\frac{\mu^2}{\beta})^2 e^{2\rho-2\phi} =0.
\label{eq8}
\end{eqnarray}
The dilaton and matter equations are
\begin{equation}
(-4\partial_+\partial_-\phi +4\partial_+\phi\partial_-\phi +
2\partial_+\partial_-\rho)e^{-2\phi}
+\left(\lambda^2 -\frac{\beta}{16}(f^2-\frac{\mu^2}{\beta})^2 \right)
 e^{2\rho-2\phi}=0,
\label{eq9}
\end{equation}
\begin{equation}
\partial_+\partial_-f +\frac{1}{4}(\beta f^3-\mu^2 f) e^{2(\rho-\phi)}=0.
\label{eq10}
\end{equation}
Adding the equations (\ref{eq8}) and (\ref{eq9}) we get
 $\partial_+\partial_-(\rho-\phi)=0$, and by making use of the subconformal
 gauge freedom we can let
\begin{equation}
\rho = \phi. \label{eq11}
\end{equation}
Now the field equations are much simplified as
\begin{eqnarray}
\partial_+^2(e^{-2\phi}) + \frac{1}{2}(\partial_+f)^2 &=&0, \label{eq12} \\
\partial_-^2(e^{-2\phi})+ \frac{1}{2}(\partial_-f)^2 &=&0,   \label{eq13} \\
\partial_+\partial_-(e^{-2\phi})+\lambda^2-\frac{\beta}{16}(f^2-\frac{\mu^2}
{\beta})^2 &=&0,
\label{eq14} \\
\partial_+\partial_-f + \frac{1}{4}(\beta f^3-\mu^2 f) &=&0. \label{eq15}
\end{eqnarray}

In the rest of this paper we will consider solutions of the above equations
 in connection with solitons and their influence upon geometry. Notice
 that the last equation (\ref{eq15}) is just a nonlinear Klein-Gordon
 equation whose solution is well known.
\vspace{1cm}

\section{Blackhole Formation by a Kink}
\hspace{0.5cm}
We begin with the soliton solution of the Klein-Gordon equation with
 a quartic ${\rm interaction^7}$:
\begin{equation}
f(x,t) = \sqrt\frac{\mu^2}{\beta}~\tanh\left[\frac{\mu \gamma}{\sqrt{2}}
((x-x_0)+v(t-t_0))\right],  \label{eq16}
\end{equation}
where
\begin{equation}
\gamma=\frac{1}{\sqrt{1-v^2}}, \label{eq17}
\end{equation}
and  $(x_0,t_0)$ is the center of the soliton. This is a traveling wave of
kink type with velocity $-v$. It is helpful to rewrite the soliton solution
 in terms of $x^\pm$ as
\begin{equation}
f(x^+,x^-) = \sqrt\frac{\mu^2}{\beta} ~\tanh{\left[\frac{\mu}{2 \sqrt 2}
(\alpha(x^+ - x_0^+)-\frac{1}{\alpha}(x^- - x_0^-))\right]} \label{eq18}
\end{equation}
where
\begin{equation}
\alpha= \sqrt{\frac{1+v}{1-v}}. \label{eq19}
\end{equation}

In order to study black hole formation by a kink we solve the equations
 (\ref{eq12})-(\ref{eq14}) with $f$ given by (\ref{eq18}). The equation
 (\ref{eq14}) is explicitly given as
\begin{equation}
\partial_+\partial_- (e^{-2\phi}) = -\lambda^2 + \frac{\mu^4}{16\beta}
{}~\mbox{sech}^4 \left[\frac{\mu}{2 \sqrt 2}(\alpha(x^+ - x_0^+)-
\frac{1}{\alpha}(x^- - x_0^-))\right], \label{eq20}
\end{equation}
which can be integrated as
\begin{equation}
e^{-2\phi} = a(x^+) +b(x^-) -\lambda^2 x^+ x^- -\frac{\mu^2}{3\beta}
 \left[\log[{\cosh({\Delta - \Delta_0})}]+\frac{1}{4}
 \tanh^2({\Delta - \Delta_0}) \right],\label{eq21}
\end{equation}
where
\begin{equation}
\Delta=\frac{\mu}{2 \sqrt 2}(\alpha ~x^+ - \frac{1}{\alpha} ~x^-).
\label{eq22}
\end{equation}
The two functions $a(x^+)$ and $b(x^-)$ are determined by the constraint
 equations (\ref{eq12}) and (\ref{eq13}). They are simply
\begin{equation}
\frac{d^2 a(x^+)}{{dx^+}^2} = 0,~~\frac{d^2 b(x^-)}{{dx^-}^2} = 0,
\label{eq23}
\end{equation}
which determines the dilaton field up to constants as
\begin{equation}
e^{-2\phi} = C + a x^+ + b x^- -\lambda^2 x^+ x^- -\frac{\mu^2}{3\beta}
 \left[\log[{\cosh({\Delta - \Delta_0})}]+\frac{1}{4}
 \tanh^2({\Delta - \Delta_0}) \right],\label{eq24}
\end{equation}
where $a$, $b$, $C$ are constants. By choosing the origin of
the coordinates suitably we can let $a=b=0$. Before we start analyzing
the geometry we summarize the solution of kink type as the followings:
\begin{eqnarray}
f(x^+,x^-)& =& \sqrt\frac{\mu^2}{\beta}~\tanh [{\Delta - \Delta_0}],
\label{eq25}\\
e^{-2\rho} =e^{-2\phi}&=&C -\lambda^2 x^+x^- -\frac{\mu^2}{3\beta}
 \left[\log[{\cosh({\Delta - \Delta_0})}]+\frac{1}{4}
 \tanh^2({\Delta - \Delta_0}) \right], \label{eq26}\\
\Delta &=&\frac{\mu}{2 \sqrt 2}(\alpha ~x^+ - \frac{1}{\alpha} ~x^-)
=\frac{\mu \gamma}{\sqrt 2}(x+vt).\label{eq27}
\end{eqnarray}

In order to see the nature of the geometry induced by the soliton it is
 convenient to divide the spacetime into three regions : $\Delta-\Delta_0
 \ll -1$, $\Delta-\Delta_0 \simeq 0$, and $\Delta-\Delta_0 \gg 1$. The
 first region ($\Delta-\Delta_0 \ll -1$)  is the area which has not yet
 been influenced by the soliton, and so we expect it to be a vacuum. We
 can indeed check that the equation (\ref{eq26}) becomes in this region
\begin{equation}
e^{-2\rho}\simeq -\lambda^2~(x^+ + \frac{\mu^3}{6\sqrt 2 \alpha\beta
 \lambda^2})~(x^- - \frac{\alpha \mu^3}{6\sqrt 2 \beta \lambda^2})\label{eq28}
\end{equation}
where the constant $C$ is taken as
\begin{equation}
C=\frac{\mu^2}{3\beta}(\Delta_0 + \frac{\mu^4}{24\beta \lambda^2}
 - \log 2 +\frac{1}{4}). \label{eq29}
\end{equation}
This clearly exhibits that the metric is that of the linear dilaton
 vacuum when the soliton has not arrived to affect the spacetime.
 In the third region ($\Delta-\Delta_0 \gg 1$), we expect the metric of a
 black hole, which can be checked by retaining only the first order term
 in (\ref{eq26}) as
\begin{equation}
e^{-2\rho}\simeq \frac{\mu^2}{3\beta} (2\Delta_0) -\lambda^2
{}~(x^+ - \frac{\mu^3}{6\sqrt 2 \alpha\beta \lambda^2})~(x^- + \frac{\alpha
\mu^3}{6\sqrt 2 \beta \lambda^2}). \label{eq30}
\end{equation}
This represents the geometry of a black hole of mass
$\lambda\frac{\mu^2}{3\beta} (2\Delta_0)$ after
 shifting $x^+$ by $\frac{\mu^3}{6\sqrt 2 \alpha\beta \lambda^2}$,
 and $x^-$ by $-\frac{\alpha \mu^3}{6\sqrt 2 \beta \lambda^2}$.
 The first region of the linear dilaton vacuum and the third one of a black
 hole are smoothly joined across the soliton wave. At near the center of the
 soliton ($\Delta-\Delta_0 \simeq 0$) we see that
\begin{equation}
e^{-2\rho}\simeq \frac{\mu^2}{3\beta}
(\Delta_0+\frac{\mu^4}{24\beta\lambda^2}-\log2+\frac{1}{4})
-\lambda^2 ~x^+ x^-  \label{eq31}
\end{equation}
which represents again the geometry of a black hole but with a different
 of mass $ \lambda\frac{\mu^2}{3\beta}
(\Delta_0+\frac{\mu^4}{24\beta\lambda^2}-\log2+\frac{1}{4})$.

The position of the apparent horizon is given by $\partial_+ \phi=0$ which is
\begin{equation}
x^- + \frac{\alpha \mu^3}{6\sqrt 2 \beta \lambda^2}~\tanh{(\Delta-\Delta_0)}
[1+\frac{1}{2}~\mbox{sech}^2({\Delta-\Delta_0})]=0.\label{eq32}
\end{equation}
At the center of the soliton wave $(\Delta-\Delta_0 =0)$  it is simply $x^-
= 0$, and at the third region $(\Delta-\Delta_0 \gg 1)$  it is $x^-
\simeq -\frac{\alpha \mu^3}{6\sqrt 2 \beta \lambda^2}$ which coincides with
 the event horizon of the black hole.
\vspace{1cm}

\section{Discussion}
\hspace{0.5cm}
It is instructive to compare our results with those of the CGHS $^1$
 where a black hole is formed by a shock wave of massless scalar fields
 traveling in the $x^-$ direction with magnitude $A$ described by the stress
 tensor
\begin{equation}
\frac{1}{2}(\partial_+ f)^2=A \delta(x^+ -x_0^+).\label{eq33}
\end{equation}
The black hole geometry caused by this shock wave is described by
 the metric
\begin{equation}
e^{-2\rho}=-A (x^+ -x_0^+) \Theta(x^+ -x_0^+)-\lambda^2 x^+ x^-,
\label{eq34}
\end{equation}
where $\Theta$ is a step function. In our case we take the relativistic
 limit $~(v\rightarrow 1)$ of the soliton. From the soliton solution
 (\ref{eq25}) we see that
\begin{eqnarray}
\frac{1}{2}(\partial_+ f)^2&=&\frac{1}{2}\left(\frac{\alpha\mu ^2}
{2\sqrt{2\beta}}\right)^2~\mbox{sech}^4 \left[\frac{\mu}{2 \sqrt 2}
(\alpha(x^+ - x_0^+)-\frac{1}{\alpha}(x^- - x_0^-))\right]\nonumber\\
&\rightarrow&\frac{\mu^3}{3\beta\sqrt{1-v}}\delta(x^+ -x_0^+).\label{eq35}
\end{eqnarray}
Therefore the magnitude of the stress tensor of a ultrarelativistic
 soliton is
\begin{equation}
A_{soliton}=\frac{\mu^3}{3\beta\sqrt{1-v}}.\label{eq36}
\end{equation}
On the other hand the metric caused by the soliton becomes
\begin{eqnarray}
e^{-2\rho} \rightarrow & &\left\{\begin{array}{ll}
 \displaystyle-\lambda^2 x^+(x^- -\frac{\mu^3}{6\beta \lambda^2\sqrt{1-v}}),
 &  ~~~ x^+-x_0^+<0, \\
 \displaystyle-\lambda^2 x^+(x^- +\frac{\mu^3}{6\beta \lambda^2\sqrt{1-v}})
+ \frac{\mu^3}{3\beta\sqrt{1-v}} ~x_0^+, & ~~~  x^+ -x_0^+>0,
\end{array} \right.
\label{eq37}\\
& &= -\lambda^2 x^+(x^- -\frac{\mu^3}{6\beta \lambda^2\sqrt{1-v}})
- \frac{\mu^3}{3\beta\sqrt{1-v}}(x^+- x_0^+)\Theta(x^+- x_0^+),
\label{eq38}
\end{eqnarray}
which again coincides with the metric of CGHS case with
\begin{equation}
A=\frac{\mu^3}{3\beta\sqrt{1-v}}.\label{eq39}
\end{equation}

In this paper we limited our consideration to classical problems.
 It is natural to investigate quantum nature of the theory such as
 Hawking radiation, conformal anomaly, and semiclassical analysis
 which require considerable amount of work and not a straightforward
 extension of previous results on Hawking radiation because of
 simutaneous presence of black hole and soliton. We plan
 to deal with these topics in a separate article.
\vspace{1cm}

\noindent
{\Large{\bf Acknowledgements}}\vspace{0.5cm} \\
This work was supported in part by the Center for Theoretical Physics (SNU),
and by the Basic Science Research Institute Program,  Ministry of Education,
1994, Project No. BSRI-94-2418, and also by S.N.U. Daewoo Research Fund.
\vspace{1cm}

\noindent
{\Large{\bf References}}
\vspace{0.5cm}

\begin{description}
\item{1.} ~C. G. Callan, Jr., S. B. Giddings, J. A. Harvey and A. Strominger,
	{\em Phys. Rev.} {\bf D45}, R1005 (1992).
\item{2.} ~S. W. Hawking, {\em Phys. Rev. Lett.} {\bf 69} 406 (1992);
	B. Birnir, S. B. Giddings, J. A. Harvey and A. Strominger,
	{\em Phys. Rev.}  {\bf D46} 638 (1992);
	T. Banks, A. Dabholkar, M.R. Douglas and M. O' Loughlin,
	{\em Phys. Rev.}  {\bf D45} 3607 (1992);
	J.G. Russo, L. Susskind and L. Thorlacius, {\em Phys. Lett.}
{\bf B292}
	13 (1992); D. A. Lowe, {\em Phys. Rev.} {\bf D47}, 2446 (1993);
	T. Piran and A. Strominger, {\em Phys. Rev.} {\bf D48}, 4729 (1993).
\item{3.} ~J. G. Russo, L. Susskind, and L. Thorlacius, {\em Phys. Rev.}
	{\bf D46} 3444 (1992); {\bf D47} 533 (1993).
\item{4.} ~A. Mikovic, {\em Phys. Lett.}  {\bf B291},  19 (1992);
       J. Navarro-Salas, M. Navarro and V. Aldaya, {\em Nucl. Phys.}
	{\bf B403}, 291 (1993).
\item{5.} ~H. S. Shin and K. S. Soh, to appear in {\em Phys. Rev.} {\bf D}
(1995).
\item{6.} ~B. Stoetzel, {\em Two-dimensional Gravitation and Sine-Gorgon
        Solitons}, {\em gr-qc}, {\bf 9501033}.
\item{7.} ~M. J. Albowitz, D. J. Kaup, A. C. Newell and H. Segur,
	{\em Phys. Rev. Lett.} {\bf 30} 1262 (1973); R. Dodd, J. Eilbeck,
	J. Gibbon, and H. Morris, {\em Solitons and Nonlinear Wave Equations}
       (Academic Press, London, 1982).
\end{description}

\end{document}